# Swap Bribery

Edith Elkind  
University of Southampton, UK and  
Division of Mathematical Sciences,  
Nanyang Technological University,  
Singapore

Piotr Faliszewski*  
Department of Computer Science  
AGH Univ. of Science and Technology  
Kraków, Poland

Arkadii Slinko  
Deptartment of Mathematics  
University of Auckland  
New Zealand

October 26, 2018

**Abstract**

In voting theory, *bribery* is a form of manipulative behavior in which an external actor (the briber) offers to pay the voters to change their votes in order to get her preferred candidate elected. We investigate a model of bribery where the price of each vote depends on the amount of change that the voter is asked to implement. Specifically, in our model the briber can change a voter's preference list by paying for a sequence of swaps of consecutive candidates. Each swap may have a different price; the price of a bribery is the sum of the prices of all swaps that it involves. We prove complexity results for this model, which we call *swap bribery*, for a broad class of election systems, including variants of approval and $k$-approval, Borda, Copeland, and maximin.

## 1 Introduction

There is a range of situations in social choice where an external actor may alter some of the already submitted votes, or the votes that the voters intend to submit. For example, a candidate can attempt to change the voters' preferences by running a campaign, which may be targeted at a particular group of voters. A more extreme (and illegal) version of this strategy involves paying voters to change their votes, or bribing election officials to get access to already submitted ballots in order to modify them. Alternatively, one can assume that the submitted votes can be contaminated with random mistakes, and a central authority should be allowed to correct the votes (preferably, by changing them as little as possible) to reveal the true winner. Indeed, this scenario is, in fact, one of the original motivations behind Dodgson's voting rule. (See papers [15, 6] for a discussion of this idea.)

All of these activities can be interpreted as changing the voters' preferences subject to a budget constraint, and can therefore be studied using the notion of bribery in elections introduced by

*Supported by AGH University of Science and Technology Grant no. 11.11.120.777.

Faliszewski, Hemaspaandra, and Hemaspaandra [9]. In their model of bribery, we are given an election (i.e., a set of candidates and a list of votes), a preferred candidate $p$, a price of each vote, and a budget $B$. We ask if there is a way to pick a group of voters whose total price is at most $B$ so that via changing their votes we can make $p$ a winner.

In the model of Faliszewski, Hemaspaandra, and Hemaspaandra [9] each voter may have a different price, but this price is fixed and does not depend on the nature of the requested change: upon paying a voter, the briber can modify her vote in any way. While there are natural scenarios captured by this model, it fails to express the fact that voters may be more willing to make a small change to their vote (e.g., swap their 2nd and 3rd most favorite candidates) than to change it completely. To account for such settings, Faliszewski [8] proposed a new notion of bribery, which he called *nonuniform bribery*. Under nonuniform bribery, a voter's price may depend on the nature of changes she is asked to implement. A similar notion called *microbribery* was considered in [10]. However, none of these papers considers the standard model of elections, in which votes are preference orders over the set of candidates. Specifically, Faliszewski [8] focused on the so-called utility-based voting, while Faliszewski et al. [10] used the irrational voter model, in which voters' preferences may contain cycles.

The goal of this paper is to study a notion of nonuniform bribery that can be used within the standard model of elections. Our model, which we call *swap bribery*, is a direct specialization of the notion of microbribery to the case of rational voters. In addition, it is also inspired by Dodgson voting rule (see Fellows, Rosamond, and Slinko [11] for a related discussion). We use the name "swap bribery" as it precisely captures the nature of our model. In swap bribery, the briber can ask a voter to perform a sequence of swaps; each swap changes the relative order of two candidates that are currently adjacent in this voter's preference list. For example, if a voter prefers $a$ to $b$ and $b$ to $c$ (we write this as $a \succ b \succ c$), she can be asked to swap $a$ and $b$, then $a$ and $c$, then $b$ and $c$, resulting in the vote $c \succ b \succ a$. Each swap has an associated price, and the total price is simply the sum of the prices of individual swaps. When preferences are viewed as orderings, a swap of adjacent candidates is a natural "atomic" operation on a vote. Moreover, one can transform any vote into any other vote by a sequence of such swaps. Hence, attaching prices to such operations provides a good model for nonuniform bribery in the standard setting.

We also study a special case of swap bribery, which we call *shift bribery*. Under this model of bribery the only allowable swaps are the ones that involve the preferred candidate. Thus, in effect, a shift bribery amounts to asking a voter to move the preferred candidate up by a certain number of positions in her preference order. As argued above, bribery can be used to model a legal approach to influencing elections, namely, campaigning: the "briber" simply invests money into trying to convince a particular group of voters that one candidate is better than another. The message and costs of the campaign can vary from one group of voters to another, which is captured by different bribery prices. In this context, shift bribery corresponds to campaigning for the preferred candidate (as opposed to discussing relative merits of other candidates), and is therefore particularly appealing.

After introducing our model of bribery, we proceed to study it from the algorithmic perspective. Our goal here is threefold. First, as argued above, despite its negative connotations, bribery may correspond to perfectly legal and even desirable behavior, and therefore we are interested in developing efficient algorithms that a potential "briber" (that is, a campaign manager) can use. Second, from a more technical perspective, we would like to pinpoint the source of computational hardness in nonuniform bribery. Indeed, when the number of candidates is unbounded, the general bribery of Faliszewski, Hemaspaandra, and Hemaspaandra [9] appears to be hard for all but the simplest election systems. In contrast, there is a number of polynomial-time algorithms for nonuniform bribery in non-standard models, such as utility-based voting or irrational voters. We would like to know whether these easiness results are tied to the increased flexibility of pricing in nonuniform bribery, or to the increased flexibility of the alternative voter models. The results of this paper, most of which are NP-completeness results, suggest that the latter is true. Finally, our paper can be viewed within the context of "computational hardness as a barrier against manipulation" line of work, pioneered by Bartholdi, Tovey, and Trick [1]. While it has since been argued that NP-hardness might not provide sufficient protection against dishonest behavior and that more robust notions of hardness are needed (see, e.g., [20, 12, 18, 19]), identifying settings in which bribery is NP-hard is a useful first step towards finding an election system that is truly resistant to dishonest behavior.

This paper is organized as follows. After providing the necessary background in Section 2, in Section 3 we formally define our model of bribery, and prove some general results about swap bribery. Section 4 contains our results on bribery in approval voting. Then, in Section 5, we consider other popular election rules, such as Borda, Copeland, and maximin. We conclude with several directions for further research in Section 6. In Appendix A we give a discussion of swap bribery for SP-AV, a recently introduced variant of approval voting [3].

## 2 Preliminaries

**Elections.** An *election* is a pair $E = (C, V)$, where $C = \{c_1, \ldots, c_m\}$ is a set of *candidates* and $V = (v_1, \ldots, v_n)$ is a list of *voters*. Each voter $v_i$ is represented via her *preference order* $\succ_i$, which is a strict linear order over the candidates in $C$.[1] For example, given $C = \{c_1, c_2, c_3\}$ and $V = (v_1, v_2)$, we write $c_2 \succ_2 c_1 \succ_2 c_3$ to denote that the second voter, $v_2$, prefers $c_2$ to $c_1$ to $c_3$. For any $C' \subseteq C$, by writing $C'$ in a preference order we mean listing all the elements of $C'$ in an arbitrary but fixed order. Similarly, $\overleftarrow{C'}$ means listing members of $C'$ in the reverse of this fixed order.

An *election system* $\mathcal{E}$ maps an election $E = (C, V)$ to a set $W \subseteq C$ of *winners*. We assume the nonunique-winner model: all members of $\mathcal{E}(E)$ are considered to be winning. All election systems considered in this paper are point-based: they assign, via some algorithm, points to candidates, and declare as winners the ones with most points. For an election $E = (C, V)$, we denote by $\text{score}_E(c_i)$ the number of points that a candidate $c_i \in C$ receives in $E$ according to a given election system. Sometimes, to disambiguate, we will indicate in the superscript the particular election system used. We will provide the definitions of the relevant election systems as we discuss them in further sections.

**Manipulation, Possible Winners, and Bribery.** In this paper we take *manipulation* to mean unweighted constructive coalitional manipulation as defined by Conitzer, Lang, and Sandholm [5]. That is, in $\mathcal{E}$-manipulation we are given an election $E = (C, V)$, a preferred candidate $p$, and a list of "manipulative" voters $V'$, and we ask if it is possible to set the preferences of voters in $V'$ so that $p$ is an $\mathcal{E}$-winner of $(C, V \cup V')$. In the $\mathcal{E}$-possible-winner problem we are given an election $E = (C, V)$, where the voters' preference are (possibly) *partial*, i.e., are given by partial orders over $C$, and we ask if it is possible to *complete* the votes so that a given candidate $p$ is an $\mathcal{E}$-winner of the resulting election. It is not hard to see that $\mathcal{E}$-manipulation is a special case of $\mathcal{E}$-possible-winner where some votes are completely specified and some (i.e., those of the manipulative voters) are completely unspecified . The study of possible-winner problems was initiated by Konczak and Lang [14] and then continued by multiple other authors (see, e.g., Walsh's overview paper [16] and the work of Xia and Conitzer [17]). Finally, in $\mathcal{E}$-bribery [9], we are given an election $E = (C, V)$, a preferred candidate $p$, a list of voters' prices and a nonnegative integer $B$, and we ask if it is possible to modify votes at a cost of at most $B$ so that $p$ becomes an $\mathcal{E}$-winner of the resulting election. (More precisely, in [9] the term "bribery" is reserved for the case where all voters have unit prices, while the more general setting described above is called \$bribery.)

---

[1]In the context of the possible-winner problem we also allow partial orders.

**Computational Complexity.** We assume familiarity with standard notions of computational complexity such as the classes P and NP, NP-completeness, and (polynomial-time) many-one reductions. Many of our hardness proofs rely on reductions from the NP-complete problem EXACT COVER BY 3-SETS (X3C) [13].

**Definition 2.1** ([13]). *An instance $(\mathcal{B}, \mathcal{S})$ of* EXACT COVER BY 3-SETS (X3C) *is given by a ground set $\mathcal{B} = \{b_1, \ldots, b_{3K}\}$, and a family $\mathcal{S} = \{S_1, \ldots, S_M\}$ of subsets of $\mathcal{B}$, where $|S_i| = 3$ for each $i = 1, \ldots, M$. It is a "yes"-instance if there is a subfamily $\mathcal{S}' \subseteq \mathcal{S}$, $|\mathcal{S}'| = K$, such that for each $b_i \in \mathcal{B}$ there is an $S_j \in \mathcal{S}'$ such that $b_i \in S_j$, and a "no"-instance otherwise.*

## 3 Swap Bribery

In any reasonable model of nonuniform bribery, one should be able to specify the price for getting a given voter to submit any preference ordering (some of these orderings may be unacceptable to the voter, in which case the corresponding price should be set to $+\infty$). However, in elections with $m$ candidates, there are $m!$ possible votes, so listing the prices of these votes explicitly is not practical. Alternatively, one could specify the bribery prices via an oracle, i.e., via a polynomial-time algorithm that, given a voter $i$ and a preference order $\succ$, outputs the price for getting $i$ to vote according to $\succ$. However, without any restrictions on the oracle, even finding a cheapest way to affect a given vote will require exponentially many queries, and providing appropriate restrictions would be challenging.

We will now present a model of bribery that allows for easy specification of bribery prices, and yet is expressive enough to capture many interesting scenarios. Our model is based on the following idea. Intuitively, an atomic operation on a given vote is a swap of two consecutive candidates. Moreover, one can transform any vote into any other vote by a sequence of such steps. It is therefore natural to assume that the price for such transformation is reasonably well approximated by the sum of the prices of individual swaps. We now proceed to formalize this approach.

Let $E = (C, V)$ be an election, where $C = \{c_1, \ldots, c_m\}$ and $V = (v_1, \ldots v_n)$. A *swap-bribery price function* is a mapping $\pi \colon C \times C \to \mathbb{N}$, which for any ordered pair of candidates $(c_i, c_j)$ specifies the price for changing a preference order $\succ$ from $\ldots \succ c_i \succ c_j \succ \ldots$ to $\ldots \succ c_j \succ c_i \succ \ldots$. Let $(\pi_1, \ldots, \pi_n)$ be a list of swap-bribery price functions. A *unit swap* is a triple $(v_k, c_i, c_j)$. A unit swap is *admissible* if $c_i$ immediately precedes $c_j$ in $v_k$'s preference order; its price is $\pi_k(c_i, c_j)$. Executing an admissible unit swap $(v_k, c_i, c_j)$ means changing $v_k$'s preference order from $\ldots \succ c_i \succ c_j \succ \ldots$ to $\ldots \succ c_j \succ c_i \succ \ldots$.

Note that we do not allow swapping non-adjacent candidates in a single step (though, of course, such a swap could be simulated by a sequence of swaps of adjacent candidates). Indeed, such a swap would change these candidates' order relative to all candidates that appear between them in the vote.

**Definition 3.1.** *Let $\mathcal{E}$ be an election system. In $\mathcal{E}$-swap-bribery we are given an election $E = (C, V)$, where $C = \{c_1, \ldots, c_m\}$, $p = c_1$, $V = (v_1, \ldots, v_n)$, a list of voters' swap-bribery price functions $(\pi_1, \ldots, \pi_n)$, and a nonnegative integer $B$ (the budget). We ask if there exists a sequence $(s_1, \ldots, s_t)$ of unit swaps such that (1) when executed in order, each unit swap is admissible at the time of its execution, (2) executing $s_1, \ldots, s_t$ ensures that $p$ is a winner of the resulting $\mathcal{E}$-election, and (3) the sum of the prices of executing $s_1, \ldots, s_t$ is at most $B$.*

As argued above, swap bribery can be used to transform any vote into any other vote. It is natural to ask if one can efficiently compute an optimal way of doing so. It turns out that the answer to this question is "yes".

**Proposition 3.2.** *Given two votes $v_1 = c_{i_1} \succ_1 \ldots \succ_1 c_{i_m}$ and $v_2 = c_{j_1} \succ_2 \ldots \succ_2 c_{j_m}$, and a swap-bribery price function $\pi$, one can compute in polynomial time the cheapest (with respect to $\pi$) sequence of swaps converting $v_1$ into $v_2$.*

*Proof.* Set $\mathcal{I}(v_1, v_2) = \{(c_i, c_j) \mid c_i \succ_1 c_j, c_j \succ_2 c_i\}$; we say that a pair of candidates $(c_i, c_j) \in \mathcal{I}(v_1, v_2)$ is *inverted*. Clearly, to obtain $v_2$ from $v_1$, it is necessary to swap each inverted pair, so the total cost of an optimal bribery is at least $s = \sum_{(c_i, c_j) \in \mathcal{I}(v_1, v_2)} \pi(c_i, c_j)$. We will now argue that one never needs to swap a pair not in $\mathcal{I}(v_1, v_2)$, or to swap a pair in $\mathcal{I}(v_1, v_2)$ more than once; this implies that the cost of an optimal bribery is exactly $s$.

Our argument is by induction on the size of $\mathcal{I}(v_1, v_2)$. If $|\mathcal{I}(v_1, v_2)| = 0$, then $v_1 = v_2$ and the statement is obvious. Now, suppose that the statement has been proved for all $v_1', v_2'$ with $|\mathcal{I}(v_1', v_2')| < k$, and consider a pair $(v_1, v_2)$ with $|\mathcal{I}(v_1, v_2)| = k$. We claim that there is a pair of candidates $(c_i, c_j) \in \mathcal{I}(v_1, v_2)$ that is adjacent in $v_1$. Indeed, suppose otherwise, and let $(c_i, c_j)$ be a pair in $\mathcal{I}(v_1, v_2)$ that is the closest in $v_1$. By our assumption, there exists at least one $c \in C$ such that $c_i \succ_1 c \succ_1 c_j$, yet $(c_i, c) \notin \mathcal{I}(v_1, v_2)$, $(c, c_j) \notin \mathcal{I}(v_1, v_2)$. Hence, we have $c_i \succ_2 c$, $c \succ_2 c_j$, so by transitivity of $\succ_2$ we conclude $c_i \succ_2 c_j$, a contradiction with $(c_i, c_j) \in \mathcal{I}(v_1, v_2)$. Hence, $\mathcal{I}(v_1, v_2)$ always contains an adjacent pair $(c_i, c_j)$. By swapping $c_i$ and $c_j$, we obtain a vote $v_1'$ that satisfies $|\mathcal{I}(v_1', v_2)| = k - 1$. Note also that $\mathcal{I}(v_1', v_2) = \mathcal{I}(v_1, v_2) \setminus \{(c_i, c_j)\}$, as the relative order of all other candidates with respect to $c_i$ and $c_j$ did not change. Hence, we can now apply our inductive hypothesis.

The argument above suggests a simple algorithm for converting $v_1$ into $v_2$: we can repeatedly scan $v_1$ for inverted pairs whose members are currently adjacent in $v_1$, and swap the corresponding candidates. The running rime of this algorithm is bounded by $n|\mathcal{I}(v_1, v_2)| = O(n^3)$. □

Proposition 3.2 shows how to optimally convert one vote into another using swaps. We can also compute in polynomial time the cheapest way of transforming a collection of votes into any other collection of votes of the same cardinality.

**Proposition 3.3.** *Given a list of votes $V = (v_1, \ldots, v_n)$, a corresponding list of price functions $(\pi_1, \ldots, \pi_n)$, and a multiset of votes $V' = \{v_1', \ldots, v_n'\}$, one can find in polynomial time an optimal swap bribery that transforms $V$ into $V'$.*

*Proof.* Let $m'$ be the number of distinct votes in $V'$. We construct a flow network $\mathcal{N}$ with source $s$, sink $t$, $m$ vertices $x_1, \ldots, x_m$ and $m'$ vertices $y_1, \ldots, y_{m'}$. There are edges of capacity 1 and cost 0 from $s$ to each $x_i$, and edges of capacity $+\infty$ and cost 0 from each $y_j$ to $t$. Furthermore, for every pair $(x_i, y_j)$ there is an edge from $x_i$ to $y_j$ that has capacity 1 and cost $c_{ij}$, where $c_{ij}$ is the cost of transforming $v_i$ into the lexicographically $j$th distinct element of $V'$ using swap bribery; the costs $c_{ij}$ can be computed in polynomial time by Proposition 3.2. Clearly, an integer maximal flow in this network has size $m$ and corresponds to a bribery that results in the multiset of votes $V'$. Moreover, a minimum-cost maximal flow is always integer, and therefore corresponds to a minimum-cost swap bribery that produces $V'$. As a minimum-cost maximal flow in a network can be computed in polynomial time, the result follows. □

Recall that a voting rule is called *anonymous* if its outcome does not depend on the order of votes in $V$. Typical voting rules are anonymous. For such rules, Proposition 3.3 suggests a polynomial-time algorithm for finding an optimal swap bribery in the important special case where the number of candidates is fixed.

**Theorem 3.4.** *For any anonymous voting rule with a polynomial-time winner determination procedure, one can compute an optimal swap bribery in polynomial time if the number of candidates is bounded by a constant.*

*Proof.* Suppose that $|C| \leq k$, where $k$ is a given constant, and let $n = |V|$. There are at most $k!$ different votes, and hence at most $n^{k!}$ different multisets of votes of size $n$; as $k$ is a constant, this quantity is polynomial in $n$. For a given multiset of votes, we can determine if it results in our preferred candidate $p$ winning the election. For each such multiset, we can compute in polynomial time the cost of an optimal swap bribery that transforms the input list of votes into this multiset using Proposition 3.3. We can then pick the multiset that makes $p$ a winner at the smallest possible cost. Clearly, the entire procedure runs in polynomial time. □

Observe that when $|C|$ is constant, the number of different multisets of votes is polynomial in $|V|$, but the number of different lists of votes may still be superpolynomial, which is why Proposition 3.3 is phrased in terms of multisets of votes rather than lists of votes.

The next result allows us to quickly derive swap-bribery hardness results from possible-winner hardness results.

**Theorem 3.5.** *Let $\mathcal{E}$ be an election system. $\mathcal{E}$-possible-winner many-one reduces to $\mathcal{E}$-swap-bribery.*

*Proof.* The input to the $\mathcal{E}$-possible-winner problem is an election $E = (C, V)$, where the voters' orders may be partial, and a candidate $p \in C$. We give a polynomial-time algorithm that transforms an instance $(C, V)$ of $\mathcal{E}$-possible-winner problem into an instance of $\mathcal{E}$-swap-bribery in which $p$ can become a winner via swap bribery of cost 0 if and only if votes in $V$ can be completed in such a way that $p$ is a winner of a resulting election.

Our construction works as follows. First, for each (possibly) partial vote $\succ_k$ in $V$ we compute a complete vote $\succ'_k$ that agrees with $\succ_k$ on each pair of candidates that are comparable under $\succ_k$. This can easily be done via, e.g., topological sorting. For each vote $\succ'_k$ we fix the following price function $\pi_k$. For each two candidates $c_i, c_j \in C$, if $c_i$ and $c_j$ are comparable under $\succ_k$ then $\pi_k(c_i, c_j) = 1$ and otherwise $\pi_k(c_i, c_j) = 0$. We output an instance of swap bribery with budget 0, preferred candidate $p$ and an election $E'$ which is identical to $E$ except that each vote $\succ_k$ is replaced by vote $\succ'_k$ associated with price function $\pi_k$.

Clearly, this reduction runs in polynomial time. We now prove its correctness. Let us fix an index $k$ and let us consider a vote $\succ'_k$ and an arbitrary vote $\succ''_k$. We claim that $\succ'_k$ can be transformed via swap bribery to equal $\succ_k''$ at cost 0 (given price function $\pi_k$) if and only if $\succ_k''$ agrees with $\succ_k$ on all pairs of candidates comparable under $\succ_k$. By the proof of Proposition 3.2, a swap bribery that transforms $\succ_k'$ to $\succ_k''$ requires exactly swapping (in some order, each pair exactly once) each pair of candidates $c_i, c_j$ such that $c_i \succ_k' c_j$ and $c_j \succ_k'' c_i$. Clearly, the cost of these swaps is 0 if and only if $\succ_k''$ agrees with $\succ_k$ on all pairs of candidates comparable under $\succ_k$. As a result, there is a completion of the votes in $E$ that makes $p$ a winner if and only if there is a swap bribery of cost 0 that makes $p$ a winner in $E'$. □

Since $\mathcal{E}$-manipulation is a special case of $\mathcal{E}$-possible-winner, we immediately obtain the following corollary.

**Corollary 3.6.** *Let $\mathcal{E}$ be an election system. $\mathcal{E}$-manipulation many-one reduces to $\mathcal{E}$-swap-bribery.*

**Shift bribery.** In some settings, the briber may be unable to ask voters to make a swap that does not involve the preferred candidate. For example, in an election campaign investing money to support another candidate may be viewed as unethical. In such cases, the only operation available to the briber is to ask a voter to move the preferred candidate up in her preference order. We will refer to this type of bribery as *shift bribery*.

Fix an election $E = (C, V)$ with $C = \{c_1, \ldots, c_m\}$, $p = c_1$, and a voter $v \in V$ with a preference order $\succ$. Suppose that $p$ appears in the $j$th position in $\succ$. We say that a mapping $\rho : \mathbb{N} \to \mathbb{N}$ is a *shift-bribery price function* for $v$ if it satisfies (1) $\rho(0) = 0$; (2) $\rho(i) \leq \rho(i')$ for $i < i' < j$; and (3) $\rho(i) = +\infty$ for $i \geq j$. We interpret $\rho(i)$ as the price of moving $p$ up by $i$ positions in $\succ$.

**Definition 3.7.** *Let $\mathcal{E}$ be an election system. In $\mathcal{E}$-shift-bribery we are given an election $E = (C, V)$, where $C = \{c_1, \ldots, c_m\}$, $p = c_1$, and $V = (v_1, \ldots, v_n)$, a list of voters' shift-bribery price functions $(\rho_1, \ldots, \rho_n)$, and a nonnegative integer $B$ (the budget). We ask if there is a sequence $(k_1, \ldots, k_n)$ of nonnegative integers such that $\sum_{i=1}^n \rho_i(k_i) \leq B$ and bribing each voter $v_i$ to shift $p$ up by $k_i$ places ensures that $p$ is a $\mathcal{E}$-winner of the resulting election.*

It is not hard to see that $\mathcal{E}$-shift-bribery is a special case of $\mathcal{E}$-swap-bribery.

**Proposition 3.8.** *For any election system $\mathcal{E}$, any election $E = (C, V)$ given by $C = \{c_1, \ldots, c_m\}$, $p = c_1$, $V = (v_1, \ldots, v_n)$, and any list $(\rho_1, \ldots, \rho_n)$ of shift-bribery price functions for $V$, we can efficiently construct a list $(\pi_1, \ldots, \pi_n)$ of swap-bribery price functions for $V$ so that the problem of $\mathcal{E}$-shift-bribery with respect to $(\rho_1, \ldots, \rho_n)$ is equivalent to the problem of $\mathcal{E}$-swap bribery with respect to $(\pi_1, \ldots, \pi_n)$.*

*Proof.* The general idea of the proof is as follows. We are given an election $E = (C, V)$, a candidate $p = c_1$, and a shift-bribery price function $\pi_i$ for each voter $v_i$. We keep the same budget in the swap bribery problem. Now, we need to provide a swap bribery price function $\pi_i'$ for each voter. Let us fix a voter number $i$ and let us renumber the candidates in $C$ so that $v_i$'s preference order is $c_k \succ_i c_{k-1} \succ_i \cdots \succ_i c_2 \succ_i \succ_i p \succ_i \cdots$. We construct $\pi_i'$ by setting (1) $\pi_i'(p, c_2) = \pi_i(1)$, (2) for each $\ell$, $3 \leq \ell \leq k$, setting $\pi_i'(p, c_\ell) = \pi_i(\ell) - \pi_i(\ell - 1)$, and (3) setting all the remaining prices of swaps to exceed the bribery budget. A simple inductive proof shows that setting $\pi'$ in this way proves the theorem. □

The analogue of Theorem 3.5 does not seem to hold for shift bribery. Hence, unlike in the case of swap bribery, it is of interest to explore the complexity of shift bribery even when the corresponding possible-winner problem is known to be hard. Another natural question in this context is whether there are election systems for which shift bribery is strictly easier than swap bribery. As our subsequent results show, the answer to this question is "yes" (assuming $P \neq NP$).

## 4 Case Study: Approval Voting

In this section we present a nearly complete view of the complexity of swap bribery and shift bribery in $k$-approval voting. The family of $k$-approval voting rules (for various values of $k$) is a simple but interesting class of election systems, including such well-known systems as plurality and veto. In $k$-approval, a voter assigns a point to each of the top $k$ candidates on her preference list. Thus, 1-approval is simply the *plurality* rule and, for $|C| = m$, $(m-1)$-approval is the *veto* rule, where each voter votes against her least desirable candidate.

We start by showing that swap bribery is easy for both plurality and veto, but that it is hard for almost all variants of $k$-approval with fixed $k$.

**Theorem 4.1.** *Swap bribery for plurality (i.e., 1-approval) and for veto (i.e., $(m-1)$-approval) is in* P. *However, for each fixed $k$ such that $k \geq 3$, swap bribery for $k$-approval is* NP-*complete, even if all swaps have costs in the set $\{0, 1, 2\}$.*

*Proof.* We split the proof into three parts, regarding plurality, regarding veto, and regarding $k$-approval for a fixed $k$, $k \geq 3$.

**Plurality.** For each vote $v$ and each candidate $c$ it is easy to compute the minimum cost of replacing the top candidate on $v$'s preference list with $c$ (by the proof of Proposition 3.2 it is enough to keep swapping $c$ with the candidates preceding her until she is the top candidate). Then, to test whether it is possible to make our preferred candidate a winner, it is sufficient to feed these "replacement" costs together with the budget and the current votes to the polynomial-time nonuniform-bribery algorithm of Faliszewski [8].

**Veto.** The case of veto is analogous to the case of plurality and we omit it.

**k-approval, fixed k, k ≥ 3.** It is easy to see that the problem is in NP as we can simply guess the swaps to attempt, and by Proposition 3.2 there are only polynomially many swaps to guess. To show NP-hardness, we give a reduction from X3C. We will first give a construction for 3-approval, and then show how to modify it for larger values of $k$. Our input X3C instance is $(\mathcal{B}, \mathcal{S})$, where $\mathcal{B} = \{b_1, \ldots, b_{3K}\}$ and $\mathcal{S} = \{S_1, \ldots, S_M\}$. In our instance of bribery, let $C = \{p\} \cup \mathcal{B} \cup D$ be the set of candidates, where $D = \{d_1, d_2, \ldots\}$ is a set of polynomially many dummy candidates, and $p$ is the preferred candidate. For each $S_j = \{b_{i_1}, b_{i_2}, b_{i_3}\} \in \mathcal{S}$, there is a voter $v_j$ who ranks $b_{i_1}, b_{i_2}, b_{i_3}$ first, followed by $d_{3j}, d_{3j+1}, d_{3j+2}$, followed by the rest of the candidates in some order, with $p$ being the last in the list. The price function for this voter is given by $\pi_j(b_{i_3}, d_{3j}) = 1$, $\pi_j(b_{i_k}, d_{3j+\ell}) = 0$ for $k = 1, 2, 3$, $\ell = 0, 1, 2$, $(k, \ell) \neq (3, 0)$, and $\pi_j(c, c') = 2$ for any other pair $(c, c')$ of candidates. Now, if our budget for this voter is 0, we cannot change his or her vote at all. If we are willing to spend 1, we can swap $b_{i_3}$ and $d_{3j}$, and then continue to move $d_{3j}, d_{3j+1}$ and $d_{3j+2}$ into the first 3 positions. Putting any candidate that was not in the top six into the top three will cost at least 2.

Let $T$ be the largest number of points that any candidate gets from such voters. We add polynomially many votes that ensure that all candidates $b_1, \ldots, b_{3K}$ have exactly $T + 1$ points, while $p$ has exactly $T$ points. In doing so, we may utilize some fresh dummy candidates $d_{9K+1}, \ldots$; we ensure that each dummy candidate gets at most 1 point. Also, we require that $p$ is listed last in all votes that do not list him or her in the top three. The cost of swapping any pair of candidates in these new votes is 2. Finally, we set our budget to be $K$.

Clearly, buying extra points for $p$ is prohibitively expensive: $p$ would have to be moved past at least $3K - 2 > K$ candidates (we can assume that $K \geq 2$). Hence, our only chance to make $p$ a winner is to take away one point from each of $b_1, \ldots, b_{3K}$. As our budget is $K$, the only way of doing this is to bribe $K$ out of the first $M$ voters in a way that corresponds to a set cover of $\mathcal{S}$.

To adapt this construction for $k$-approval, $k > 3$, we modify all votes by adding $k - 3$ fresh dummy candidates to the top of each vote, and make it prohibitively expensive to move those candidates. □

The above theorem does not address the issue of the complexity of swap bribery in 2-approval. However, recently it was shown that the possible winner problem for 2-approval is NP-complete [2]. Thus, swap bribery for 2-approval also is NP-complete.

In contrast to Theorem 4.1, shift bribery for $k$-approval is easy for all values of $k$. Thus, shift bribery can indeed be easier than swap bribery.

**Theorem 4.2.** *Shift bribery for $k$-approval is in* P *for any $k < m$.*

The idea of the proof is that in $k$-approval shift bribery the only reasonable action that the briber has, per each voter, is to bribe that voter to move $p$ up into $k$th position (or do nothing if the voter already approves of $p$). With this observation at hand, one can apply techniques used in the proof of Theorem 4.1.

Let us consider the NP-completeness part of Theorem 4.1. There we assume that both the number of candidates and the number of voters are parts of the input (i.e., are not bounded by any fixed constant). We have seen that the first requirement is necessary: by Theorem 3.4 swap bribery becomes easy if the number of candidates is constant. It is therefore natural to ask if the number of voters plays a similar role. It turns out that if $k$ is bounded by a constant, swap bribery is easy for each fixed number of voters.

**Theorem 4.3.** *For each fixed $k$, swap bribery for $k$-approval is in* P *if the number of voters is bounded by a constant.*

*Proof.* Consider an election $E = (C, V)$, where $C = \{c_1, \ldots, c_m\}$, $V = (v_1, \ldots, v_n)$, a preferred candidate $p \in C$, and a budget $B$. (Each voter, of course, also has a swap-bribery price function.) Let $C_1, \ldots, C_T$ be the list of all $k$-element subsets of $C$; note that $T = \binom{m}{k} = \text{poly}(m)$. For a given vote $v$, we can compute the cost of moving the candidates from a given $k$-element subset $C_t$ into top $k$ positions in $v$. Indeed, suppose that $C_t = \{c_{i_1}, \ldots, c_{i_k}\}$, and $c_{i_1}$ is the first of these candidates to appear in $v$, $c_{i_2}$ is second, etc. Then this cost is simply the cost of moving $c_{i_1}$ into the top position by successively swapping it with all candidates that are above him, followed by moving $c_{i_2}$ into the second position, etc. To see why this naive algorithm is optimal, note that it only swaps pairs that are inverted in the sense of Proposition 3.2, i.e., ones that have to be swapped anyway.

We can now go over all lists of the form $(C_{i_1}, \ldots, C_{i_n})$, $i_j \in \{1, \ldots, T\}$ for $j = 1, \ldots, n$, and for each such list compute the cost of the optimal bribery that for $j = 1, \ldots, n$ transforms the $j$th input vote into a vote that lists the candidates in $C_{i_j}$ in the top $k$ positions. There are at most $\binom{m}{k}^n = \text{poly}(m)$ such lists; we accept if at least one of them has cost at most $B$ and bribing the voters to implement it ensures $p$'s victory. □

On the other hand, when $k$ is unbounded, swap bribery becomes difficult even if there is just one voter. To prove this result, we reduce from the NP-complete problem BALANCED BICLIQUE (BB) (see [13])

**Definition 4.4** ([13]). *An instance of* BB *is given by a bipartite graph $G = (U, W, E)$, where $|U| = |W| = N$ and $E \subseteq U \times W$, and a natural number $K \leq N$. We ask if there are sets $U' \subseteq U$ and $W' \subseteq W$ such that $|U'| = |W'| = K$ and $(u, w) \in E$ for all $u \in U'$, $w \in W'$.*

Intuitively, the reason why swap bribery for $k$-approval is difficult for large values of $k$ is that it may be to the benefit of the briber to move around candidates that are ranked above our preferred candidate. Doing so may allow him to then move $p$ via swaps of lower cost.

**Theorem 4.5.** *When $k$ is a part of the input, swap bribery for $k$-approval is* NP*-complete even for a single voter.*

*Proof.* It is easy to see that our problem is in NP. We focus on the NP-hardness proof. We give a reduction from BB (see Definition 4.4 above). Suppose that we are given an instance of BB with $U = \{u_1, \ldots, u_N\}$, $W = \{w_1, \ldots, w_N\}$. Our election system will have $2N+1$ candidates $u_1, \ldots, u_N$, $w_1, \ldots, w_N, p$, where $p$ is the preferred candidate, and a single voter $v$ with preference ordering $U \succ W \succ p$. The price function is given by $\pi(u_i, u_j) = 0$, $\pi(w_i, w_j) = 0$ for all $i, j = 1, \ldots, N$, $\pi(w_i, p) = 1$, $\pi(u_i, p) = 0$ for all $i = 1, \ldots, N$, $\pi(u_i, w_j) = 0$ if $(u_i, w_j) \in E$ and $\pi(u_i, w_j) = N - K + 1$ otherwise. Finally, we set $k = N + 1$ and $B = N - K$.

Suppose that we have a "yes"-instance of BB, and let $(U', W')$ be the corresponding witness. Then we can first reorder $U$ and $W$ for free so that $U \setminus U' \succ U'$, $W' \succ W \setminus W'$, then swap $U'$ and $W'$ (which is free, since $(U', W')$ is a biclique in $G$), and, finally, move $p$ past $W \setminus W'$ and $U'$, paying $|W \setminus W'| = N - K = B$.

Conversely, suppose that there is a successful bribery for $v$. Let $U'$ be the set of candidates from $U$ that end up below $p$, and let $W'$ be the set of candidates from $W$ that end up above $p$ after the bribery. Observe that this means that we had to swap each pair $(u, w) \in U' \times W'$, and hence $(u, w) \in E$ for all $(u, w) \in U' \times W'$, as otherwise we would have exceeded our budget. We had to pay 1 for swapping $p$ with each of the candidates in $W \setminus W'$, so $|W \setminus W'| \leq N - K$ and hence $W' \geq K$. On the other hand, $p$ ended up among the top $N + 1$ candidates, so $|W'| + |U \setminus U'| \leq N$, and hence $|U'| \geq K$. Pick $U'' \subseteq U'$, $W'' \subseteq W'$ so that $|U''| = |W''| = K$. The pair $(U'', W'')$ is a balanced biclique of the required size in $G$ because we have started with a successful bribery. □

**Bribery in SP-AV.** A related popular voting rule is *approval voting* (without the "$k$-" prefix), where voters can approve of (give a point to) any number of candidates. Traditionally, this rule is considered in the setting where the voters' preferences are expressed as 0/1-vectors rather than linear orders, and nonuniform bribery for approval has already been thoroughly studied [9, 8]. In Section A we provide some discussion of swap bribery in SP-AV, a variant of approval recently introduced by Brams and Sanver [3] and whose computational study was initiated by Erdélyi, Nowak, and Rothe [7].

## 5 Further Voting Rules and Shift Bribery

In this section we consider election systems other than approval, starting with Borda. In a Borda election with $m$ candidates each voter assigns to each candidate $c$ as many points as the number of candidates that the voter ranks below $c$. The possible winner problem for Borda is NP-complete [17] and thus, via Proposition 3.5 we have that Borda-swap-bribery is NP-complete. Thus, we will now focus on Borda-shift bribery.

**Theorem 5.1.** *Shift bribery for Borda is NP-complete.*

*Proof.* Clearly, shift bribery for Borda is in NP. To show completeness, we reduce from X3C. Let $(\mathcal{B}, \mathcal{S})$ be an instance of X3C where $\mathcal{B} = \{b_1, \ldots, b_{3K}\}$ and $\mathcal{S} = \{S_1, \ldots, S_M\}$ is a family of 3-subsets of $\mathcal{B}$. Set $C = \mathcal{B} \cup \{p\}$, $V = (v_1, \ldots, v_{2M+2})$. For each $S_i \in \mathcal{S}$ voter $v_i$ has preference order $S_i \succ_i p \succ_i \mathcal{B} - S_i$. Voter $v_{M+i}$ has the same order, but reversed. Voters' $v_{2M+1}$ and $v_{2M+2}$ preferences are $\mathcal{B} \succ_{2M+1} p$ and $\overleftarrow{\mathcal{B}} \succ_{2M+2} p$, respectively. Our goal is to ensure that $p$ is a winner via a bribery of cost at most $K$. The voters have the following shift-bribery price functions. For each voter $v_i$ such that $1 \leq i \leq M$, we set $\rho_i(1) = \rho_i(2) = \rho_i(3) = 1$. For each $i$ such that $1 \leq i \leq M + 2$ and for each applicable shift value $k$ we set $\rho_{M+i}(k) = K + 1$. Thus, given that our budget is $K$, the only voters that can be bribed are $v_1, \ldots, v_M$. We claim that $p$ can become a winner of this election via a bribery of cost at most $K$ if and only if $(\mathcal{B}, \mathcal{S})$ is a "yes"-instance.

It is easy to see that voters $v_1, \ldots, v_{2M}$ assign the same number of points to each candidate. Let us call this number $L$. Voters $v_{2M+1}$ and $v_{2M+2}$ assign 0 points to $p$ and $3K + 1$ points to each member of $\mathcal{B}$. Thus, in total we have $\text{score}_E(p) = L$ and for each $b_i \in \mathcal{B}$ we have $\text{score}_E(b_i) = L + 3K + 1$. It is easy to see that if $(\mathcal{B}, \mathcal{S})$ is a "yes"-instance of X3C then bribing those voters among $v_1, \ldots v_M$ that correspond to a cover to rank $p$ first costs $K$ and ensures $p$'s victory. This is so, because there are exactly $K$ voters to bribe and bribing them has the following effect: $p$'s score increases by $3K$ (bribing each one of them increases $p$'s score by 3) and the score of each $b_i \in \mathcal{B}$ decreases by 1. In effect, all candidates tie as winners. The reverse direction holds via a simple argument. □

On the positive side, there exists a polynomial-time 2-approximation algorithm for Borda-shift-bribery.

**Theorem 5.2.** *There exists a polynomial time algorithm that, given an instance $I$ of shift bribery with a preferred candidate $p$, outputs a sequence of shifts that makes $p$ a Borda winner, and whose cost is at most $2c$, where $c$ is the cost of an optimal Borda-shift bribery for $I$.*

*Proof.* Fix an instance $I$ of Borda-shift bribery. Suppose that in $I$ the optimal shift bribery has cost $c$ and moves $p$ up by $k$ positions in total.

It is easy to see that in $I$ *any* bribery that shifts $p$ up by at least $2k$ positions results in $p$ being a winner. This is so because in the optimal solution shifting $p$ up by $k$ positions increases $p$'s score by $k$ and decreases every other candidate's score by at most $k$. Thus, altogether $p$ gets at most $2k$ points compared to each of the other candidates. We obtain the same effect by shifting $p$ up by $2k$ positions.

Now, suppose that we know $k$. Let $B$ be the cheapest bribery that shifts $p$ up by $k$ positions. This bribery can be computed by a dynamic programming algorithm as follows. For each $i = 1, \ldots, n$ and $k' = 1, \ldots, k$, let $f(i, k')$ be the cost of the cheapest shift bribery that moves $p$ up by $k'$ positions in the preferences of the first $i$ voters. We have $f(1, k') = \rho_i(k')$ for $k' \leq m - k_1$, where $k_1$ is the position of $p$ in the first vote, and $f(1, k') = +\infty$ for $k' > m - k_1$. Further, we have $f(i+1, k') = \min\{f(i, k' - k'') + \rho_{i+1}(k'') \mid k'' = 1, \ldots, m - k_{i+1}\}$, where $k_{i+1}$ is the position of $p$ in the $(i+1)$st vote. Obviously, the cost of $B$ is given by $f(n, k)$, and one can compute $B$ itself using standard techniques. Observe that the cost of $B$ is at most $c$.

Now, $B$ includes some $j$ shifts, $j \leq k$, that also appear in the optimal solution. Suppose that we know the value of $j$. Let us imagine that we first execute these $j$ shifts. After doing so, we get an instance $I'$ that still allows the remaining $k - j$ shifts of the optimal solution. Thus, given $I'$, one can find $k - j$ shifts that ensure $p$'s victory and so, by the observation in the previous paragraph, any $2(k - j)$ shifts from $I'$ suffice to make $p$ a winner. Let $I''$ be the instance obtained after executing $B$. Clearly, one can transform $I'$ into $I''$ using $k - j$ shifts. Therefore, in $I''$ any bribery that shifts $p$ by $k - j$ positions results in $p$ winning. Thus, after executing $B$, we pick the cheapest bribery $B'$ that shifts $p$ up by $k - j$ positions. These $k - j$ shifts cost at most $c$, because there are the $k - j$ unused shifts from the optimal solution, whose cost is at most $c$. As a result, we ensure $p$'s victory via $2k - j$ shifts, and pay at most $2c$.

Now, this algorithm of course assumes knowing $k$ and $j$. When solving an arbitrary instance, we do not know them, but we can try all combinations. □

Quite interestingly, the randomized approximation algorithm of Caragiannis et al. [4] designed to compute Dodgson scores can be used (with very minor changes only) to solve Borda-shift-bribery. Caragiannis et al.'s algorithm takes as input an election $(C, V)$ and a candidate $p$ and outputs an approximate number of shifts of $p$ that guarantee that $p$ is a Condorcet winner (i.e., the algorithm returns the approximate Dodgson score of $p$). However, internally, for each candidate $c$ the algorithms simply stores a minimal number of times that $p$ needs to pass $c$ on some voter's preference list. The algorithm finds a sequence of shifts that guarantees that each candidate is passed at least the required number of times (these numbers are selected in a way that ensures that $p$ is a Condorcet winner, but the correctness of the algorithm does not depend on this fact; we can use arbitrary numbers instead). In terms of Borda, this means that the algorithm internally specifies by how many points each candidate's score should be decreased via shifting $p$. Thus, we can use the algorithm of Caragiannis et al. to find an approximately optimal shift bribery that guarantees that each candidate other than $p$ has at most some prespecified number of points (we should try all possible numbers of points; there are only polynomially many of them). If this already makes $p$ a winner then we have an approximate solution. If that does not yet make $p$ a winner then we simply need to shift $p$ up until she is a winner and our approximate solution's cost is the cost computed by the randomized algorithm + the cost of this final shifting. (Technically, the algorithm of Caragiannis et al. treats all shifts as having unit cost, but it can quite easily be adapted to consider arbitrary prices.)

Nonetheless, our 2-approximation result is much stronger. The approximation ratio of the algorithm of Caragiannis et al. [4] is $\log m$, where $m$ is the number of candidates. (However, for the case of Dodgson, Caragiannis et al. [4] show that their result is optimal, assuming $P \neq NP$.)

We now turn to elections defined via considering majority contests between pairs of candidates. Specifically, we consider maximin and Copeland$^\alpha$, where $\alpha$ is a rational number, $0 \leq \alpha \leq 1$. Given an election $E = (C, V)$ where $C = \{c_1, \ldots c_m\}$ and $V = (v_1, \ldots v_n)$, we define $N_E(c_i, c_j) = |\{v_k \mid c_i \succ_k c_j\}|$. Let $\alpha$ be a rational number such that $0 \leq \alpha \leq 1$. Copeland$^\alpha$ score of a candidate $c_i$, score$_E^\alpha(c_i)$, is defined as score$_E^\alpha(c_i) = |\{c_j \mid N_E(c_i, c_j) > N_E(c_j, c_i)\}| + \alpha |\{c_j \mid N_E(c_i, c_j) = N_E(c_j, c_i)\}|$. That is, $c_i$'s score is the number of candidates that he or she defeats in head-to-head majority contests plus $\alpha$ times the number of candidates with whom $c_i$ ties such contests. Maximin score of a candidate $c_i$, score$_E^m(c_i)$, is $\min_{i \neq j} N_E(c_i, c_j)$.

**Theorem 5.3.** *Shift bribery is* NP-*complete for maximin and, for each rational $\alpha$ between 0 and 1, for Copeland$^\alpha$.*

|   | $p$ | $t$ | $c$ | $\mathcal{B}$ |
|---|---|---|---|---|
| $p$ | $-$ | $L$ | $L+K$ | $L+K-1$ |
| $t$ | $L+2K$ | $-$ | $K$ | $2L+2K$ |
| $c$ | $L+K$ | $2L+K$ | $-$ | $2L+2K-1$ |
| $\mathcal{B}$ | $L+K+1$ | $0$ | $1$ | $\leq 2L+2K$ |

Table 1: Results of head-to-head contexts between candidates in the election constructed in the proof of Theorem 5.3. An entry in row $a$ and column $b$ is the value $N_E(a,b) - M$ (or its upper bound). Entries regarding $\mathcal{B}$ give the value for each member of $\mathcal{B}$.

*Proof.* Let us first consider Copeland voting. A close inspection of the proof that bribery is NP-complete for each rational value of $\alpha$, $0 \leq \alpha \leq 1$, given in [10] shows that that proof works, upon specifying proper price functions for shift bribery as well.[2] To avoid repetition, we skip the proof.

Let us now move to maximin. As usual, it is easy to see that the problem is in NP and we focus an a reduction that shows its NP-completeness. We will reduce X3C to shift bribery for maximin.

Let $(\mathcal{B}, \mathcal{S})$ be an input instance of X3C where $\mathcal{B} = \{b_1, \ldots, b_{3K}\}$ and $\mathcal{S} = \{S_1, \ldots, S_M\}$. Let $L$ be a nonnegative integer whose exact value we will specify later. We form an election $E = (C, V)$, where $C = \{p, t, c\} \cup B$ and where $V$ contains $2(M + K + L)$ voters, $v_1, \ldots v_{2(M+K+L)}$. Voters are divided into two main groups. Voters $v_1, \ldots v_{2M}$ implement the structure of our input X3C problem and the remaining voters create appropriate padding, so that the scores of the candidates in $C$ require the briber to solve the X3C instance we reduce from.

For each nonnegative integer $i$, $1 \leq i \leq M$, voter $v_i$ has preference order $t \succ_i S_i \succ_i p \succ_i \mathcal{B} - S_i \succ_i c$ and voter $v_{M+i}$ has preference order that is the reverse of $v_i$'s. Thus, $N_{(C,(v_1,\ldots,v_{2M}))}$ is a constant function that equals $M$ for each pair of candidates. We set the remaining $2K + 2L$ candidates as follows:

1. There are $L$ voters with preference order $p \succ c \succ t \succ \mathcal{B}$.

2. There are $K - 1$ voters with preference order $t \succ p \succ c \succ \mathcal{B}$.

3. There is a single voter with preference order $t \succ \mathcal{B} \succ p \succ c$.

4. There are $L + K$ voters with preferernce order $c \succ t \succ \mathcal{B} \succ p$.

We set the budget $B = K$. For each $v_i$, $1 \leq i \leq M$, $v_i$'s shift-bribery function is such that $\pi_i(0) = 0$ and $\pi_i(1) = \pi_i(2) = \pi_i(3) = \pi_i(4) = 1$. For each of the remaining voters we set shift-bribery prices that do not allow to change their votes at cost $K$ or lower. That is, within the budget, the briber can only choose to shift $p$ by 1, 2, 3, or 4 positions forward on the preference lists of some voters among $v_1, \ldots, v_M$. The cost of such a bribery is equal to the number of those voters that the briber chooses to affect. We claim that it is possible to ensure that $p$ is a winner via such a bribery if and only if $(\mathcal{B}, \mathcal{S})$ is a "yes"-instance of X3C.

We will first show that if it is possible to ensure $p$'s victory via a shift bribery of cost at most $K$ then $(\mathcal{B}, \mathcal{S})$ is a "yes"-instance. Table 1 shows the values of $N_E(\cdot, \cdot) - M$ for each pair of candidates in $C$. The table contains entries "minus $M$," because voters $v_1, \ldots, v_{2M}$ contribute $M$ points to each candidate's score. All the candidates in $\mathcal{B} \subset C$ have similar scores and so both the column and the row labeled $\mathcal{B}$ applies to each candidate in $\mathcal{B}$. The value that lies at the intersection of this row and this column, cell $(\mathcal{B}, \mathcal{B})$ so to say, should be interpreted as an inequality that holds for each $b_i, b_j \in \mathcal{B}$. That is, for each $b_i, b_j \in \mathcal{B}$ we have $N_E(b_i, b_j) - M \leq 2L + 2K$.

---

[2]In essence, we need to use the same trick as, e.g., in the proof of Theorem 5.1. That is, we set the shift-bribery prices so that shifting $p$ by one position was as useful as shifting it to the top of a voter's ranking.

We set $L = K$ (though, we encourage the reader to think of $L$ as "a large value"). Via trivial calculation, we see that the candidates have the following scores: $\text{score}_E^m(p) = M + L$, $\text{score}_E^m(t) = M + K$, $\text{score}_E^m(c) = M + K + L$, and for each $b_i \in \mathcal{B}$, $\text{score}_E^m(b_i) \leq M$. Thus, before any bribery attempts, $c$ is the winner of the election. Also, since the only votes that we can affect via shift bribery are $v_1, \ldots, v_M$, it is impossible to lower $c$'s score. Thus, via inspection of the entries in the $p$-row of Table 1 it is easy to see that to ensure $p$'s victory in the election, we have to increase $N_E(p, t)$ by at least $K$ and increase each of $N_E(p, b_i)$ by at least 1. However, this is possible only if we choose to shift-bribe exactly $K$ voters among $v_1, \ldots, v_M$ to rank $p$ first (so that $N_E(p, t)$ increases by $K$) and if those voters correspond to a cover of $\mathcal{B}$ (so that each $N_E(p, b_i)$ increases by exactly 1). Thus, if it is possible to ensure $p$'s victory via a shift bribery of cost at most $K$, then $(\mathcal{B}, \mathcal{S})$ is a "yes"-instance of X3C. It is easy to see that the other direction of the reduction holds as well and so the proof is complete. □

It is interesting to compare the results of this section with those of [10], which shows that for irrational voters microbribery for Copeland$^0$ and for Copeland$^1$ is in P. In fact, we can also show that microbribery for the case of irrational voters is also in P for Borda and maximin (though we omit these results due to limited space and our focus on rational voters).

This is a further (meta)-argument that perhaps the main source of hardness in many election problems stems from the necessity of dealing with preference orders rather than from the design of particular election systems.

## 6 Conclusions

We have introduced a notion of swap bribery and its cousin shift bribery, and analyzed their complexity in several well-known election systems such as plurality, $k$-approval, Borda, Copeland, and maximin. It turns out that, in sharp contrast to the easiness results for microbribery [10] and nonuniform bribery in utility-based systems [8], swap bribery is NP-hard for many of these systems. This is quite surprising as our swap bribery is essentially the microbribery model adapted to the rational-voter setting.

Our work leads to several open problems. First, it would be useful to identify natural special cases of our setting for which one can find an optimal swap bribery in polynomial time. Another approach to tackling computational hardness is constructing effective approximation algorithms for swap bribery and shift bribery. Theorem 5.2 makes the first step in this direction; designing approximation algorithms for other voting rules and for swap bribery is a topic for future research.

## A  Swap Bribery in SP-AV

In this section we consider variant of approval voting recently introduced by Brams and Sanver [3] called *sincere-strategy preference-based approval voting (SP-AV)*. In SP-AV each voter $v_i$, in effect, provides a preference order and an integer $\ell_i$, $1 \leq \ell_i \leq m-1$, indicating how many of her top-ranked candidates she approves of. Erdélyi, Nowak, and Rothe [7] initiated computational study of SP-AV by considering *control* in SP-AV.

In defining bribery for SP-AV, it is natural to allow the briber to ask the voters both to swap adjacent candidates (as in swap bribery) and to change the number of candidates they approve of. We formalize this idea as follows. A *mixed bribery problem* for SP-AV is given by an election $E = (C, V)$, $|V| = n$, with a preferred candidate $p$, a budget $B$, a list $(\pi_1, \ldots, \pi_n)$ of swap bribery price functions, and a list $(\sigma_1, \ldots, \sigma_n)$ of *approval threshold price functions*, where $\sigma_i : \mathbb{Z} \to \mathbb{N}$ satisfies $\sigma(0) = 0$. We interpret $\sigma_i(k)$ as the price for changing the number of candidates that the $i$th voter approves of by $k$; note that we allow $k < 0$.

Theorems 4.1 and 4.5 immediately imply that mixed bribery is hard even if the briber is not allowed to change the approval thresholds, i.e. $\sigma_i(k) = +\infty$ for all $v_i \in V$, $k \in \mathbb{Z}$, as long as either (1) $\ell_i \geq 3$ for all $v_i \in V$ or (2) $\ell_i = n/2 + 1$ for at least one $v_i \in V$. Interestingly, it is also hard if the briber is *only* allowed to change the approval thresholds and even if the corresponding prices are linear.

**Theorem A.1.** *Mixed bribery for SP-AV is NP-complete even if $\pi_i(c_j, c_k) = +\infty$ for all $v_i \in V$ and all $c_j, c_k \in C$ and $\sigma_i(k) = |k|$ for all $v_i \in V$.*

*Proof.* It is easy to see that the problem is in NP. To show NP-completeness we now give a reduction from X3C. Let $(\mathcal{B}, \mathcal{S})$ be our input instance, where $\mathcal{B} = \{b_1, \ldots, b_{3K}\}$ and $\mathcal{S} = \{S_1, \ldots, S_M\}$. Let $T = \{t_1, \ldots, t_M\}$ be a set of padding candidates. We form an SP-AV election $E = (C, V)$, where $C = \mathcal{B} \cup T \cup \{p, e\}$ and with $M + 2K + 2$ voters $(v_1, \ldots v_{M+2K+2})$. The first $M$ voters correspond to the sets in $\mathcal{S}$. For each $i$, $1 \leq i \leq M$, $v_i$ has preference order $t_i \succ_i S_i \succ_i p \succ_i \mathcal{B} - S_i \succ_i T - \{t_i\} \succ_i e$ and approves only of the topmost candidate, i.e., $\ell_i = 1$. The remaining voters report the following preference orders and approval counts:

1. $K + 1$ voters report $e \succ \mathcal{B} \succ T \succ p$ and approve only of the topmost candidate, $e$,

2. 1 voter reports $p \succ \mathcal{B} \succ T \succ e$ and approves only of the topmost candidate, $p$,

3. $K$ voters report $\mathcal{B} \succ T \succ e \succ p$ and approve of all the members of $\mathcal{B}$.

For each of the voters we use price function $\pi$ that does not allow any swaps, and a price function $\sigma$ such that for each integer $k$, $\sigma(k) = |k|$ (here $|k|$ means the absolute value of $k$).

We claim that $p$ can become a winner of this election via mixed bribery of cost at most $B = 3K$ if and only if $(\mathcal{B}, \mathcal{S})$ is a "yes"-instance of X3C. Via a routine calculation we see that in this election candidates have the following scores:

1. $\mathrm{score}_E(p) = 1$,

2. $\text{score}_E(e) = K + 1$,

3. $\text{score}_E(b_i) = K$,

4. $\text{score}_E(t_i) = 1$.

Due to SP-AV rules (every voter has to approve of at least one candidate and no voter can approve of all candidates), it is easy to see that it is impossible to decrease $e$'s score via mixed bribery. Thus, the only way to ensure that $p$ is a winner is to increase $p$'s score by $K$ without increasing the score of any of the $b_i$'s by more than 1. The only way to increase $p$'s score is via bribing some of the voters $v_1, \ldots v_M$. However, bribing them to increase $p$'s score costs 3 for each of the voters. Thus, a bribery that increases $p$'s score by $K$ has to include bribing $K$ of the voters $v_1, \ldots v_M$ to increase their approval counts by 3. If such a bribery is to ensure $p$'s victory, those bribed voters in $v_1, \ldots v_M$ have to correspond to a cover of $B$ via sets from $\mathcal{S}$ because increasing $p$'s score by 1 via bribing some voter $v_i$, $1 \leq i \leq M$, also means that the score of all members of $S_i$ increases by 1. Thus, if it is possible to ensure $p$'s victory via count bribery then $(\mathcal{B}, \mathcal{S})$ is a "yes"-instance. On the other hand, it is easy to see that if $(\mathcal{B}, \mathcal{S})$ is a "yes"-instance then it is possible to ensure $p$'s victory via a mixed bribery of cost at most $3K$. □